\documentclass[12pt]{article}
\usepackage{latexsym,epsfig,graphicx,epstopdf,amsmath,amssymb,amscd,undertilde,multirow,chicago,psfrag,paralist,dsfont,url, listings, booktabs, algorithm, algorithmicx, algpseudocode, amssymb, amsmath, bm}

\usepackage[skip=5pt]{caption}
\usepackage[titletoc]{appendix}
\usepackage[T1]{fontenc}
\usepackage[american]{babel}
\usepackage{comment}
\usepackage{varwidth}
\usepackage{verbatim}

\usepackage{helvet}

\usepackage{listings}
\usepackage{comment}
\usepackage{hyperref}

\newcommand{\minitab}[2][l]{\begin{tabular}{#1}#2\end{tabular}}

\begin{document}

\title{StatLLM: A Dataset for Evaluating the Performance of Large Language Models in Statistical Analysis}

\author{
Xinyi Song, Lina Lee, Kexin Xie, Xueying Liu, Xinwei Deng, and Yili Hong\footnote{Corresponding Author. Email: yilihong@vt.edu}\\[1.5ex]
{Department of Statistics, Virginia Tech, Blacksburg, VA 24061}
}

\date{}

\maketitle

\begin{abstract}
The coding capabilities of large language models (LLMs) have opened up new opportunities for automatic statistical analysis in machine learning and data science. However, before their widespread adoption, it is crucial to assess the accuracy of code generated by LLMs. A major challenge in this evaluation lies in the absence of a benchmark dataset for statistical code (e.g., SAS and R). To fill in this gap, this paper introduces StatLLM, an open-source dataset for evaluating the performance of LLMs in statistical analysis. The StatLLM dataset comprises three key components: statistical analysis tasks, LLM-generated SAS code, and human evaluation scores. The first component includes statistical analysis tasks spanning a variety of analyses and datasets, providing problem descriptions, dataset details, and human-verified SAS code. The second component features SAS code generated by ChatGPT 3.5, ChatGPT 4.0, and Llama 3.1 for those tasks. The third component contains evaluation scores from human experts in assessing the correctness, effectiveness, readability, executability, and output accuracy of the LLM-generated code. We also illustrate the unique potential of the established benchmark dataset for (1)  evaluating and enhancing natural language processing metrics, (2) assessing and improving LLM performance in statistical coding,  and (3) developing and testing of next-generation statistical software -- advancements that are crucial for data science and machine learning research.


\end{abstract}

\section{Introduction}
\subsection{Background and Motivation}

Statistical analysis plays an important role in machine learning and data science.  The programming capabilities of large language models (LLMs) have unlocked new possibilities for automatic statistical analysis. Current versions of LLMs (e.g., ChatGPT, \shortciteNP{openai2024gpt4technicalreport}, and Llama, \shortciteNP{touvron2023llamaopenefficientfoundation}) can generate statistical code in SAS (\citeNP{sassoftware}) and R (\citeNP{R}), two widely used languages for statistical analysis in academia and industry. However, before these models see widespread adoption, it is essential to evaluate the accuracy and reliability of the code they generate.

One of the key challenges in evaluating the performance of LLMs for statistical programming is the lack of a systematically designed benchmark dataset, particularly for SAS and R. Unlike fields such as natural language processing (NLP) and computer vision, where standardized datasets are widely available for model evaluation, statistical programming lacks such resources. This gap hinders the systematic assessment of LLM-generated statistical code, making it difficult to quantify performance in terms of correctness, executability, and output quality. Thus, there is a pressing need to develop a comprehensive benchmark dataset tailored for statistical code evaluation.

To bridge this data gap, we introduce StatLLM, an open-source dataset specifically designed to evaluate the performance of LLMs in statistical analysis. The StatLLM dataset is structured into three key components: (i) statistical analysis tasks, (ii) LLM-generated SAS code, and (iii) expert human evaluation scores, which are detailed in Section~\ref{sec:statllm.dataset}. Below, we provide an overview of each component and its contribution.

The first component, statistical analysis tasks, comprises a diverse collection of problem descriptions covering a broad range of statistical topics, including data visualization, descriptive statistics, hypothesis testing, regression and ANOVA, generalized linear models, survival analysis, model selection, and nonparametric statistics. These topics are commonly taught at the upper undergraduate and master's levels in statistics programs. Each task is accompanied by detailed dataset descriptions and human-verified SAS code, ensuring a rigorous evaluation of LLMs' ability to interpret statistical requirements and generate accurate statistical code.

The second component, LLM-generated SAS code, consists of code produced by three LLMs: ChatGPT 3.5, ChatGPT 4.0, and Llama 3.1 70B. For each model, we provide a problem description and dataset details, prompting it to generate SAS code for the statistical analysis. SAS is selected due to its structured syntax, which allows most statistical analyses to be performed within ten lines of code, making human evaluation more efficient and manageable.

The third component, expert human evaluation scores, offers a rigorous assessment of LLM-generated code based on multiple criteria, including code quality, executability, and output accuracy. These evaluations are conducted by statistics experts, requiring substantial human effort and domain expertise. This human-centered evaluation approach serves as a gold standard, setting StatLLM apart from other benchmarking datasets that often rely solely on automated metrics.

In addition to detailing the data collection and organization process, we highlight the unique potential of the StatLLM dataset to advance research in various areas of artificial intelligence (AI) and data science. A comprehensive discussion is provided in Section~\ref{sec:illustration.impact}, but we offer a brief overview below. First, StatLLM serves as a valuable resource for evaluating and improving NLP metrics used to assess statistical code. It enables researchers to assess the suitability of existing NLP metrics for evaluating statistical code and facilitates the development of specialized metrics tailored to the unique requirements of statistical analysis. These tailored metrics can provide a more reliable and domain-aware assessment of LLMs' performance in generating statistically sound and executable code. Second, the human rating scores in StatLLM provide a comprehensive assessment of LLMs' capabilities in statistical programming. This evaluation helps identify weaknesses in LLM-generated code, providing actionable insights to improve the quality and reliability of LLMs in statistical programming. Third, StatLLM establishes a unique foundation for developing and testing next-generation statistical software that enables users to interact with statistical tools using natural language. It allows researchers to systematically evaluate how well LLMs can perform end-to-end data analysis tasks. These advancements are particularly significant for data science and machine learning research, where the demand for reliable, efficient, and interpretable analytical tools continues to grow.

\subsection{Related Work}

For AI technology to gain widespread acceptance, its reliability and performance must be rigorously tested (\shortciteNP{hong2023statistical}). Several studies have evaluated LLMs' programming capabilities. \shortciteN{atkinson2023chatgpt} analyzed GPT 3.5's performance across multiple programming languages. \shortciteN{bubeck2023sparks} examined GPT 4.0's performance on 40 coding problems, showing it outperforms GPT 3.5 in code generation. Similarly, \shortciteN{yeadon2024comparison} compared the coding performance of human students, GPT 3.5, and GPT 4.0 on university-level programming tasks. \shortciteN{Songetal2025-coderating} conducted human evaluations of LLMs' performance in SAS programming across various statistical tasks.

Benchmarks have been developed to assess LLMs' programming abilities, primarily for widely used languages like Python. \shortciteN{chen2021evaluatinglargelanguagemodels} introduced HumanEval, a benchmark for evaluating LLMs on Python function-level code generation. Other notable benchmarks include MBPP (\shortciteNP{austin2021program}), APPS (\shortciteNP{hendrycks2021measuring}), Multi-HumanEval (\shortciteNP{athiwaratkun2022multi}), and CoderEval (\shortciteNP{yu2024codereval}). The ``BigCodeBench'' study (\shortciteNP{zhuo2024bigcodebenchbenchmarkingcodegeneration}) highlights the need for evaluating LLMs on diverse function calls and complex instructions to enhance real-world applicability.

Class-level benchmarks, such as ClassEval (\shortciteNP{du2024evaluating}), provide a more realistic assessment by examining how functions interact within a class. Multilingual benchmarks like MBXP and Multi-HumanEval1 have attempted to address language diversity by translating Python into other languages (\shortciteNP{athiwaratkun2022multi}). CoderEval (\shortciteNP{yu2024codereval}) further advanced real-world testing by incorporating non-standalone functions. RepoCoder introduced a repository-level framework for code completion, incorporating context-aware evaluation through the RepoEval benchmark (\shortciteNP{zhang2023repocoder}). ConvCodeWorld's multi-turn interaction paradigm (\shortciteNP{han2025convcodeworld}) better simulated iterative statistical workflows, while CodeEditorBench (\shortciteNP{guo2024codeeditorbench}) focused on code editing tasks, revealing LLMs' challenges in preserving context during incremental modifications. Despite these advancements, benchmarks for statistical languages such as SAS and R remain underrepresented.

Establishing data repository is crucial for research at the intersection of AI and statistics (\shortciteNP{Zhengetal2025-datareview}). Several efforts have focused on developing benchmarks for statistical analysis. \shortciteN{hu2024infiagent} introduced InfiAgent-DABench, a benchmark for assessing LLMs' data analysis capabilities in Python. \shortciteN{zhu2024largelanguagemodelsgood} presented StatQA, which evaluates LLM performance in elementary statistical analysis. \shortciteN{liu2024llmscapabledatabasedstatistical} proposed a benchmark to assess LLMs' ability to perform statistical and causal analysis with real-world data. \shortciteN{huang2024evaluating} compared ChatGPT 4.0's data analysis capabilities to traditional statistical software. Unlike these studies, StatLLM offers a novel contribution by providing a comprehensive dataset specifically designed to evaluate LLMs' proficiency in statistical programming, with a primary focus on SAS.

\begin{figure}
\begin{center}
\includegraphics[width=0.95\textwidth]{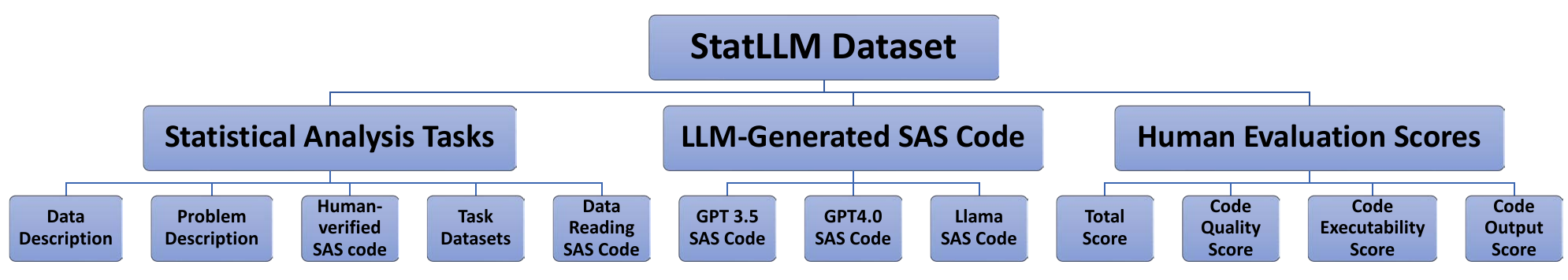}
\caption{An overview of the structure of the StatLLM dataset.}\label{fig:data.structure}
\end{center}
\end{figure}

\subsection{Overview}
The remainder of the paper is organized as follows. Section~\ref{sec:statllm.dataset} introduces the StatLLM datasets. Section~\ref{sec:illustration.impact} illustrates their use and impact on statistical analysis and LLM research. Section~\ref{sec:ethics.fairness} discusses ethics and fairness. Finally, Section~\ref{sec:conclusions} presents the conclusions.

\section{The StatLLM Dataset}\label{sec:statllm.dataset}

\subsection{The Structure of StatLLM}
This section provides an overview of the structure of StatLLM, as illustrated in Figure~\ref{fig:data.structure}. StatLLM consists of three key components: statistical analysis tasks, LLM-generated SAS code, and human-verified SAS code, which are detailed in Sections~\ref{sec:stat.ana.tasks}, \ref{sec:llm.sas.code}, and \ref{sec:human.score.rating}, respectively. The dataset is available in the GitHub repository: \url{https://github.com/yili-hong/StatLLM}, with a description of its folder structure provided in Appendix~\ref{sec:online.data.repo}.

\subsection{Statistical Analysis Tasks}\label{sec:stat.ana.tasks}

A key objective of the StatLLM is to develop a comprehensive set of statistical analysis tasks to evaluate LLM capabilities. During the 2022-2023 academic year, researchers assembled a diverse collection of datasets from various public online resources. The data collection spanned multiple disciplines, including biological sciences, medical research, engineering, and social sciences. The final compilation consists of 65 CSV datasets with detailed descriptions, which are used to create 207 distinct statistical analysis tasks. Each problem includes both a problem description and its implementation in SAS programming language. This material is systematically organized for analysis purposes.

Each data description is documented with key aspects like dataset names, contextual background, variable specifications, and comprehensive variable descriptions.

\noindent\textbf{Example of data description}:

{\sffamily \footnotesize 
Measurements were made on men involved in a physical fitness course at N. C. State Univ. The variables are Age (years), Weight (kg), Oxygen intake rate (ml per kg body weight per minute), time to run 1.5 miles (minutes), heart rate while resting, heart rate while running (same time Oxygen rate measured), and maximum heart rate recorded while running. The dataset is named as fitness. The independent variables are Age, Weight, RunTime, RestPulse, RunPulse and MaxPulse. The dependent variable is Oxygen.
}

Problem descriptions are crafted to clearly communicate the statistical requirements to the LLMs. To ensure reliability, researchers manually verifies and tests each SAS code solution, confirming that it produced accurate results for its corresponding problem.

\noindent\textbf{Example of problem description}:

{\sffamily \footnotesize 
For the fitness data, we are interested in doing the FORWARD model selection methods for the response variable Oxygen. Display only the selection summary table.
}

\noindent\textbf{Example of human-verified SAS code}:

{\sffamily \footnotesize 
\noindent proc reg data = fitness plots=(criteria sbc);\\
model Oxygen=Age Weight RunTime RunPulse\\
RestPulse MaxPulse / selection=forward\\
details=summary; run;
}

In summary, the 65 datasets span various domains, including education, medical research, business analytics, finance, and sports statistics. Each dataset is briefly described, highlighting key variables and focus areas. For example, educational datasets include school performance metrics, student assessments, and psychological factors influencing academic success. Medical and health-related datasets cover patient recovery, blood pressure readings, and treatment efficacy studies. Business and sales analytics datasets feature company sales records, employee surveys, and inventory data. Financial datasets track historical market trends, while sports datasets focus on player performance statistics. For each dataset, we provide a SAS data step code, enabling users to load the data into SAS for analysis.

The statistical analysis tasks are designed to be solvable by students with statistical training at the undergraduate to master's level. These tasks cover a broad range of analytical techniques, including basic data analysis, visualization, regression, ANOVA, hypothesis testing, generalized linear models (GLM), and advanced methods such as survival analysis and nonparametric statistics. Table~\ref{tab:statistical.analysis.tasks} summarizes the topics covered across the 207 tasks, highlighting the comprehensive scope of statistical techniques included.

\begin{table}
\begin{center}
\caption{Summary of topics included in those 207 statistical analysis tasks.}\label{tab:statistical.analysis.tasks}
{\small
\begin{tabular}{c|l}\hline\hline
Category & Example Tasks \\\hline
\minitab[c]{Data\\ Management}&
\minitab[l]{Dataset information; Sorting;
Subsetting; Variable creation;\\
 Data transformation.} \\\hline
\minitab[c]{Descriptive\\ Statistics}&
\minitab[l]{Basic statistics; Quantiles;
Winsorized mean; \\ Contingency tables;
Group comparisons; Missing value.}\\\hline
Correlation&
\minitab[l]{Pearson correlation matrices;
categorical variable associations.}\\\hline
\minitab[c]{Data \\ Visualization}&
\minitab[l]{Scatterplot; Boxplot;
Histograms; Probability plots;\\
Time series plots; Bar-charts.}\\\hline
\minitab[c]{Hypothesis\\ Testing \\Confidence \\ Intervals}&
\minitab[l]{One-sample tests; Two-sample tests;
 Multiple comparisons; \\ Turkey's test;
 Fisher's LSD; Paired t-tests;
 Proportion t-tests; \\ Welch's t-test
 Normality test; Chi-square test.}\\\hline
ANOVA &
\minitab[l]{One-way and two-way ANOVA;
Interaction; \\ Multiple comparisons;
Treatment effects; CRD.}\\\hline
Regression&
\minitab[l]{Simple, multiple, quadratic
 regression; Robust regression; \\
 Iterated reweighted least squares;
 Model diagnostics; \\ Influence; Cook's D;
 Collinearity diagnostics; VIF analysis; \\
 Homogeneity of Variance; Lack of fit;
 Variable transformations.}\\\hline
GLM&
\minitab[l]{Logistic regression, Poisson regression;
Mixed effects.}\\\hline
\minitab[c]{Model \\Selection}&
\minitab[l]{AIC-based; Forward;
 Backward; MAXR method.}\\\hline
\minitab[c]{Advanced \\ Methods}&
\minitab[l]{Survival analysis; Cox Model;
Non-parametric statistics.}\\\hline
\hline

\end{tabular}
}
\end{center}
\end{table}

\begin{figure}
\begin{center}
\includegraphics[width=0.8\textwidth]{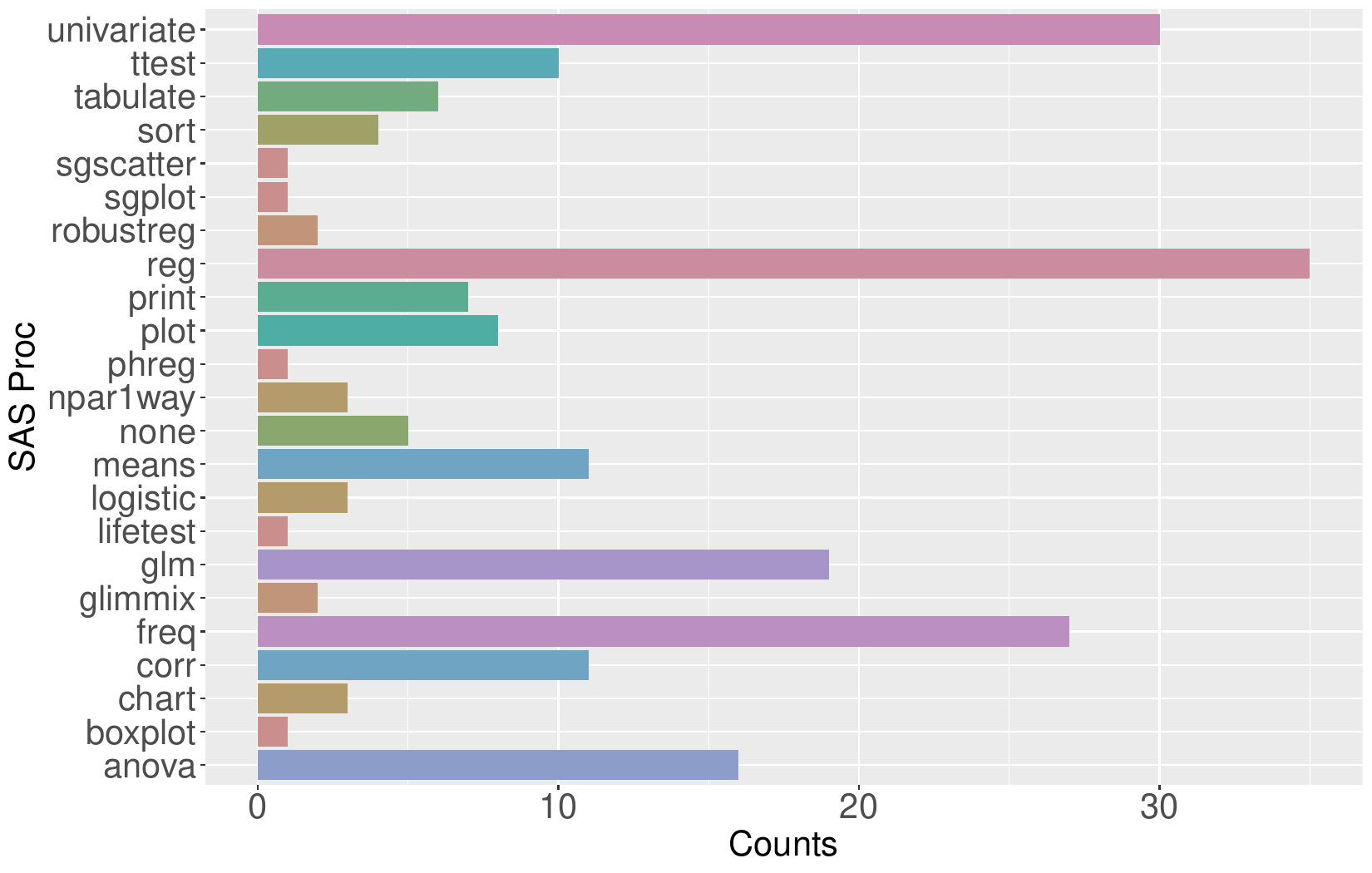}
\caption{Visualization of SAS PROCs used in the statistical analysis and its corresponding frequency. }\label{freq:procs}
\end{center}
\end{figure}

To solve these 207 statistical analysis tasks, a variety of SAS PROCs are used, including t-tests, regression, logistic regression, correlation analysis, and various plotting functions. Figure~\ref{freq:procs} visualizes those SAS PROCs used in the statistical analysis and its corresponding frequency. SAS code solutions are generally concise, with most SAS code implementations requiring fewer than 10 lines.

\subsection{LLM-Generated SAS Code}\label{sec:llm.sas.code}

We first give a brief description of the three LLMs chosen for the study.

\subsubsection{GPT 3.5 and GPT 4.0}

ChatGPT (\shortciteNP{openai2024gpt4technicalreport}) is an advanced AI model by OpenAI, built on a generative pre-trained transformer. It processes large-scale data to enable natural-language interactions, assisting with queries, problem-solving, research, and programming tasks.

ChatGPT has evolved across versions, improving problem-solving, efficiency, and adaptability in coding. GPT 3.5, released in March 2022, features a transformer-based architecture with dense attention layers, supporting code generation and debugging, particularly for beginners (\shortciteNP{shirafuji2023exploringrobustnesslargelanguage}). However, its performance declines in complex, multi-step programming tasks. \shortciteN{sherman2024} showed that GPT 3.5 struggled with non-compliant C code, frequently missing obvious errors while focusing on minor issues, highlighting its limitations in complex code analysis.

Released on March 14, 2023, GPT 4.0 significantly improves comprehension, reasoning, and accuracy over previous versions (\shortciteNP{openai2024gpt4technicalreport}). Its refined architecture enhances complex problem-solving, code generation, and debugging across languages. In this study, we use both GPT 3.5 and GPT 4.0 to generate SAS code.

\subsubsection{Llama}
The Llama series, developed by Meta AI, includes open-source LLMs for NLP tasks like code generation. Llama 1 debuted in February 2023, followed by the fully open-source Llama 2 in July 2023. Code Llama (\shortciteNP{bap2024codellamaopenfoundation}), built on Llama 2, rivals proprietary models like OpenAI's Codex on benchmarks such as HumanEval and MBPP. However, it struggles with complex contexts, prompt dependency (\shortciteNP{chen2021evaluatinglargelanguagemodels}), and security risks (\shortciteNP{e25060888}). Llama 3 enhances multilingual understanding, coding, reasoning, and tool use (\shortciteNP{dubey2024llama3herdmodels}). Available in configurations like Llama 3.1 8B, 70B, and 405B, it scales from 8 to 405 billion parameters, impacting code generation quality. The 8B model suits moderate tasks, the 70B model excels in complex analysis, and the 405B model handles advanced, multi-task programming. In this study, we used Llama 3.1 70B to generate SAS code.

\subsubsection{SAS Code Generation}

For the three selected LLMs, we utilize their APIs to generate SAS code for statistical analysis tasks. Each model receives a data description and problem statement, prompting it to produce SAS code as a solution. The generated code is then extracted from the LLM output, filtered to remove irrelevant content, and stored for future evaluation and assessment.

\subsection{Human Evaluation Scores}\label{sec:human.score.rating}

\begin{figure}
\begin{center}
\includegraphics[width=0.95\textwidth]{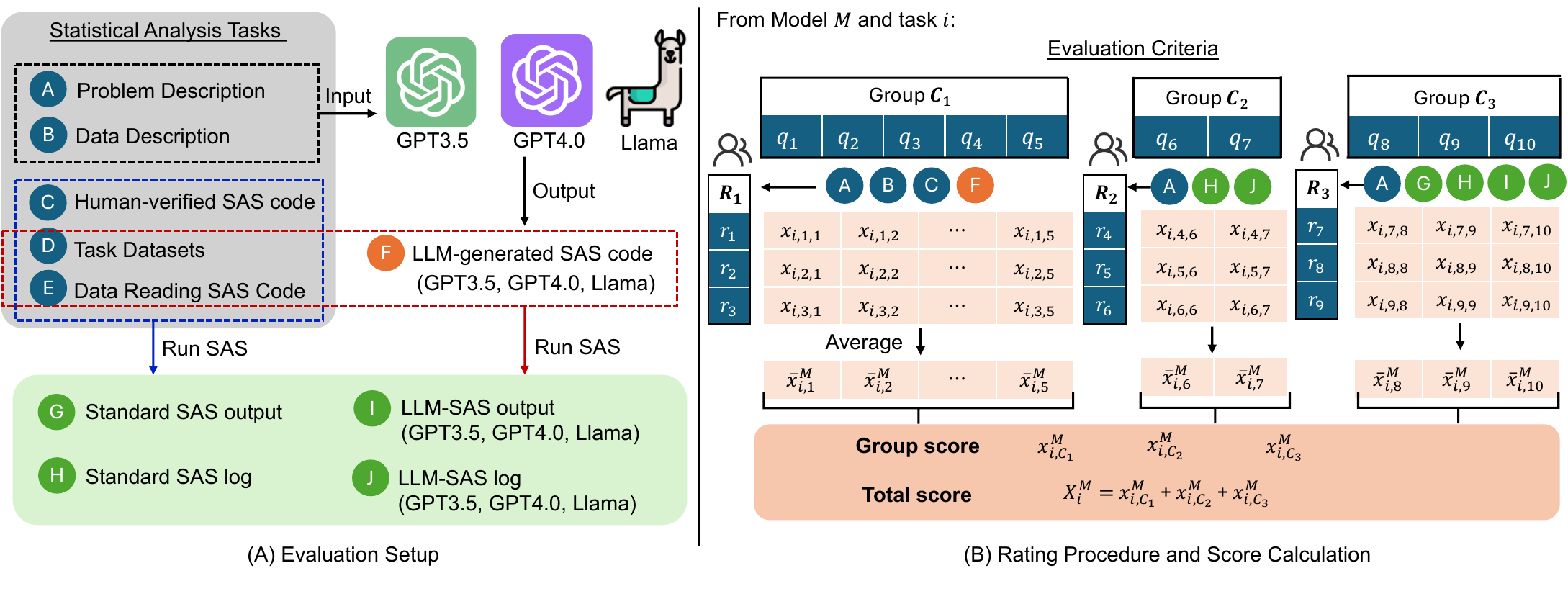}
\caption{Flowchart illustrates the main steps in human evaluation of LLM-generated SAS code.}\label{fig:human.flowchart}
\end{center}
\end{figure}

\shortciteN{Songetal2025-coderating} conducted a comprehensive evaluation of SAS code generated by LLMs. Here, we provide a brief overview of the rating process, while the detailed 10 rating criteria can be found in \shortciteN{Songetal2025-coderating}. Figure~\ref{fig:human.flowchart} illustrates the flowchart for human evaluation. As illustrated in Figure~\ref{fig:human.flowchart}(A), the evaluation process begins with collecting statistical analysis tasks, generating SAS code using LLMs, and executing the generated SAS code to produce logs and outputs. Figure~\ref{fig:human.flowchart}(B) outlines the evaluation procedure. The human evaluation framework is structured around ten criteria $\{q_k\}_{k=1}^{10}$, grouped into three primary components $\{C_1, C_2,  C_3\}$ designed to assess distinct aspects of the generated code: Code Correctness and Readability, Executability, and Output Correctness and Quality. These aspects are rated using a standardized five-point scale, ranging from ``Strongly Disagree'' (1) to ``Strongly Agree'' (5).

The first component ($C_1=\{q_k: k=1,\dots,5\}$), Code Correctness and Readability, evaluates the technical accuracy and logical coherence of the code. It ensures that the data processing steps and model structure align with the analysis objectives, are free of errors, and are presented in a clear and organized manner. The second component ($C_2=\{q_k: k=6,7\}$), Executability, assesses whether the code runs successfully without errors or warnings. If the code executes flawlessly, it is awarded the maximum score of 10 points, bypassing further subcategory evaluations. The third component ($C_3=\{q_k: k=8,9,10\}$), Output Correctness and Quality, examines the relevance, clarity, and conciseness of the generated output in addressing the specified statistical problem. If no output is produced, all criteria in this section are assigned zero points.

The evaluation process is conducted by a panel of nine raters, denoted as $\{r_j\}_{j=1}^{9}$, selected for their expertise in SAS programming and statistical analysis. The raters are divided into three groups $\{R_1, R_2, R_3\}$, each responsible for evaluating one of the three components, and their individual scores are averaged to produce the group score for their respective component. Specifically, $R_1=\{r_j\}_{j=1}^3$ is assigned to evaluate Code Correctness and Readability ($C_1$), $R_2=\{r_j\}_{j=4}^6$ evaluates Executability ($C_2$), and $R_3=\{r_j\}_{j=7}^9$ assesses Output Correctness and Quality ($C_3$). To ensure consistency and accuracy, all raters undergo extensive training prior to the evaluation, familiarizing them with the assessment criteria and the rating scale. Additionally, each group is provided with tailored materials relevant to their assigned evaluation component, ensuring that their assessments are both precise and contextually grounded. For Group 1, focused on Code Correctness and Readability, raters receive a problem description, data description, standard SAS code, and corresponding outputs generated by three LLMs. Group 2, tasked with Executability, evaluates the SAS log files to determine whether the code runs without errors or warnings, assigning a perfect score for flawless execution. Group 3 assesses Output Correctness and Quality by reviewing the SAS log and output files to evaluate alignment with problem requirements and clarity of results.

Let $x_{i,j,k}^{M}$ denote the score assigned by rater $r_j$ to model $M$ on criterion $q_k$ within group $C_g$ for $i$th evaluation task. For each evaluation task $i$ ($i=1,\dots, N$) and a given LLM $M$, the average group score, denoted as $x_{i,g}^M$ , is computed as the mean of the scores assigned by the raters for the criteria in their respective group and is given by:
$$x_{i,g}^M=\frac{1}{3}\sum_{j \in R_g}\sum_{k \in C_g}x_{i,j,k}^{M}, \quad \text{for}\ g=1,2,3,$$
where  $R_g$ is the set of raters assigned to evaluate component $C_g$. The total score for each model, $X_i^{M}$ is then calculated as the sum of average group scores across the three groups: $X_i^{M}=\sum_{g=1}^3 x_{i,g}^M$. To alleviate potential bias during the assessment, the responses of each evaluation task from three candidate models are randomly permuted, obscuring the association between answers and specific models. Specifically, for each task $i$ ($i=1,\dots, N$), a permutation function $\pi_i : \{1, 2, 3\} \rightarrow \{1, 2, 3\}$ is applied, which specifies the position of each model's response after random shuffling. The anonymized group score presented to $i$th task for group $C_g$ of model $M$ are then expressed as: $x_{i,g}^{\pi_i(M)}$, where $\pi_i(M)$ is the permuted model index. Similarly, the permuted total score is recorded as $X_i^{\pi_i(M)}$. Finally, we calculate the group and total scores, which are available in StatLLM.

\section{Illustrations and Impact}\label{sec:illustration.impact}

In this section, we discuss and illustrate how the StatLLM dataset can be used to advance research in statistics and data science.

\subsection{NLP Metrics and Benchmarking}\label{sec:nlp.metrics.scores}
As discussed earlier, human evaluation is labor-intensive and does not scale efficiently. With both LLM-generated and human-verified SAS code available, we can leverage automatic NLP metrics to compare the two versions of code and obtain a performance rating to the LLM-generated code. In the literature, various NLP metrics have been used to evaluate the performance of AI-generated code in programming languages such as Python and C++, including BLEU (\shortciteNP{papineni-etal-2002-bleu}), ROUGE (\shortciteNP{rogue2014}), METEOR (\shortciteNP{Lavie2009TheMM}), CodeBLEU (\shortciteNP{Ren2020CodeBLEUAM}), BERTScore (\shortciteNP{zhang2020bertscoreevaluatingtextgeneration}), and CodeBERTScore (\shortciteNP{zhou-etal-2023-codebertscore}). However, the applicability of these NLP metrics to statistical coding warrants further investigation.

\begin{figure}
\begin{center}
\includegraphics[width=0.9\textwidth]{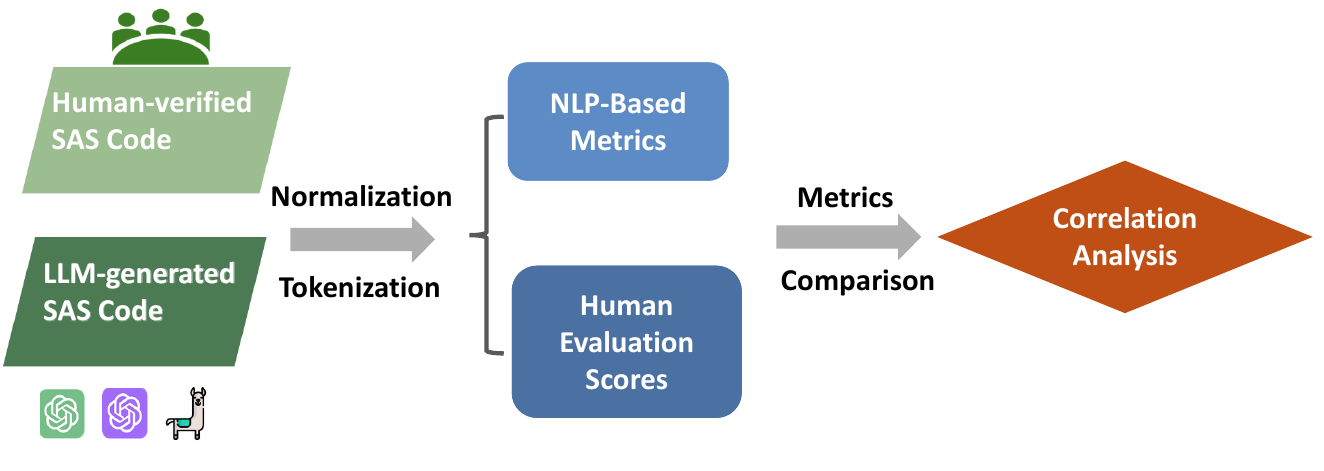}
\caption{Illustration of correlation computing for NLP metric scores vs human rating scores.}\label{fig:correatlion.cmpt.illustration}
\end{center}
\end{figure}

Note that human rating scores in StatLLM can be considered as the ground truth, we can evaluate the correlation between NLP metric scores and human ratings. Figure~\ref{fig:correatlion.cmpt.illustration} illustrates the process of computing this correlation. In this study, we consider eight NLP metrics: Meteor, BLEU, CodeBERT, ROUGE-1, ROUGE-2, ROUGE-L, Jaccard similarity, and character F-score (ChrF). For details on the computation of these metrics, refer to Section~\ref{sec:calc.nlp.metrics} in the Appendix.

Figure~\ref{fig:correlation_example GPT 4.0} presents a heatmap of Pearson correlations between NLP metric scores and human rating scores for SAS code generated by GPT 4.0. The first row, for instance, displays the correlations of the total human score with the eight NLP metric scores. The heatmap reveals that Rouge-2 and Rouge-L exhibit the highest correlations with human scores across total score, code quality score, code executability score, and output quality score. In contrast, CodeBERT demonstrates the lowest correlations with human scores.

This analysis demonstrates that the StatLLM dataset can be used to rank the effectiveness of NLP metrics in evaluating statistical code generated by LLM models. However, the results indicate that overall NLP metric scores exhibit only a moderate correlation with human scores for GPT 4.0. This suggests that there is still room for improvement in NLP metrics.

\begin{figure}
\begin{center}
\includegraphics[width=0.85\textwidth]{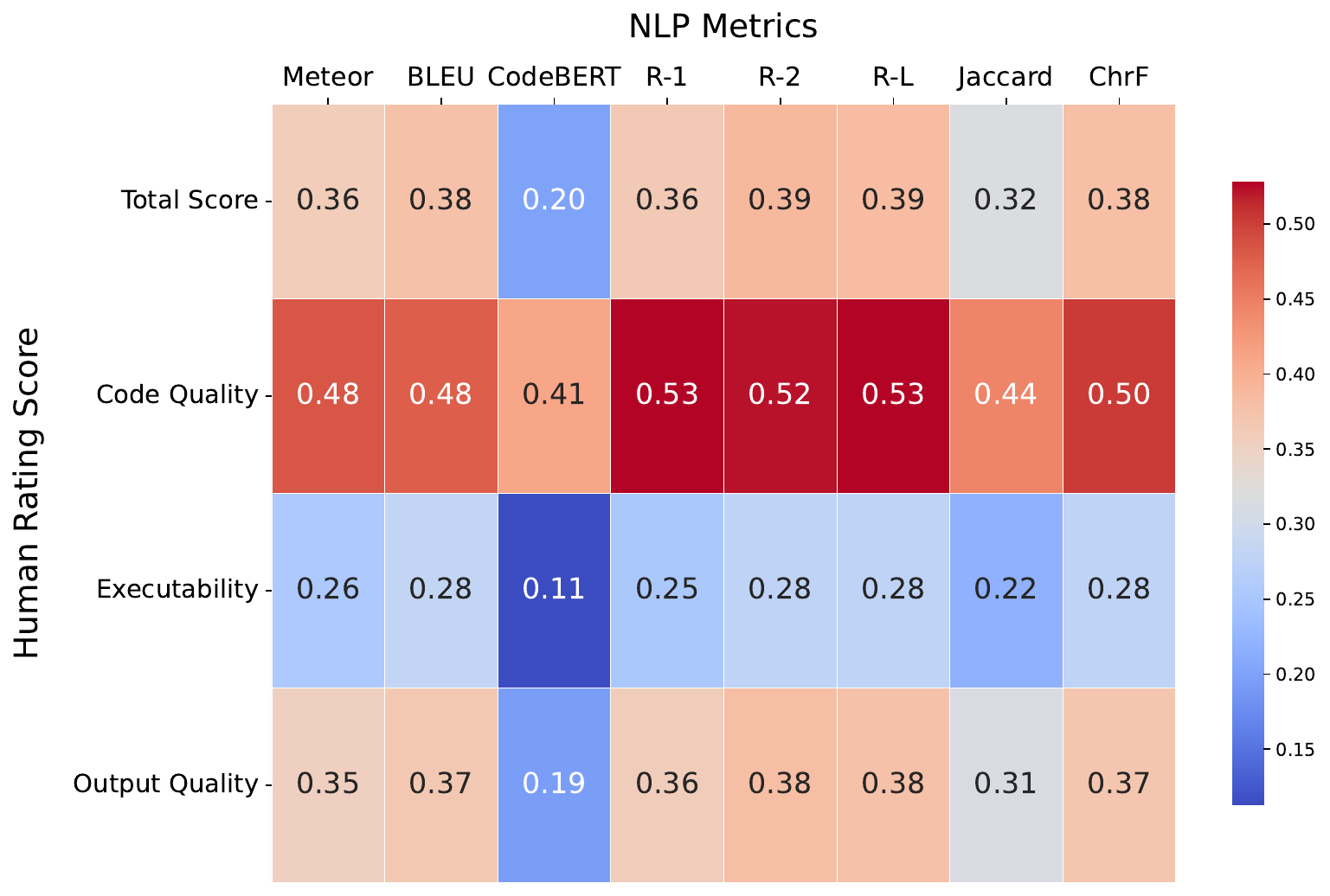}
\caption{Heatmap of correlations among NLP metric scores and human rating scores for GPT 4.0 SAS codes. Keys: R-1=ROUGE-1, R-2=ROUGE-2, and R-L=ROUGE-L.  }\label{fig:correlation_example GPT 4.0}
\end{center}
\end{figure}

\subsection{Improving NLP Metrics for Assessing Statistical Code}
Here, we illustrate one approach to leveraging the StatLLM dataset to improve NLP metrics for assessing statistical code. Specifically, we use the eight automatic NLP metric scores from Section~\ref{sec:nlp.metrics.scores} as inputs to predict the total human score. To increase sample size, we aggregate all rating scores from three LLM models, resulting in a total of 621 observations. The dataset is then randomly split, with 75\% for training and 25\% for testing.

\begin{figure}
\begin{center}
\includegraphics[width=0.5\textwidth]{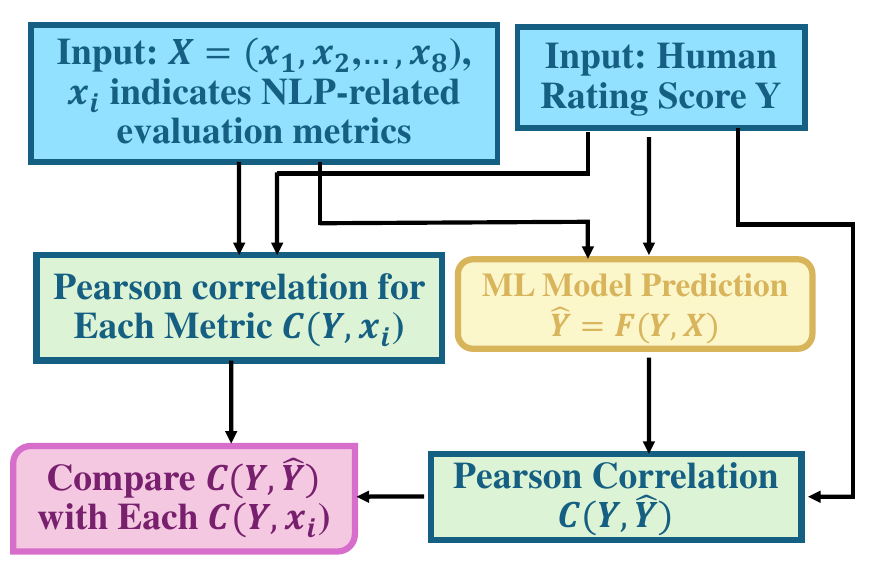}
\caption{Illustration of model training and correlation computation using machine learning methods to predict human rating scores based on NLP metric scores.}\label{fig:ML.NLP.pred.human}
\end{center}
\end{figure}

We consider four statistical and machine learning (ML) methods for prediction: linear regression model, random forest (\shortciteNP{breiman2001random}), XGBoost (\shortciteNP{chen2015xgboost}), and a deep learning model with a simple structure (\shortciteNP{Goodfellow-et-al-2016}). Figure~\ref{fig:ML.NLP.pred.human} illustrates the model training and correlation computation process using machine learning methods to predict human rating scores based on NLP metric scores. As before, Pearson correlation is used for evaluation.

Table~\ref{tab:metric.corr.test.set} presents the correlations between the predicted and true total human scores using various new metrics built with machine learning methods on the test set, as well as the correlations between true total human score and existing NLP metrics. The results show that all ML-based metrics outperform existing NLP metrics in correlation. Among them, XGBoost achieves the highest correlation of 0.434, while the best-performing existing metric, Rouge-2, has a correlation of 0.367, representing an improvement of approximately 18\%. This highlights the potential of leveraging the StatLLM dataset to enhance existing NLP metrics.

\begin{table}
\begin{center}
\caption{Correlations between the predicted and true total human scores based on various new metrics built with machine learning methods using the test set, and correlations between the true total human score and NLP metrics. Keys: Corr.=Correlation, Reg.=Regression, RF=Random Forest, XGB=XGBoost, DL=Deep Learning, C-BERT=CodeBERT.}\label{tab:metric.corr.test.set}
\begin{tabular}{cc|cc|cc}\hline\hline
\multicolumn{2}{c|}{New Metrics} &  \multicolumn{4}{c}{Existing NLP Metrics}\\\hline
Method & Corr. & Method & Corr. &Method & Corr. \\\hline
Reg. 	    &	0.382  & Meteor   &	0.243  & Rouge2	& \textbf{0.367}\\
RF   &	0.402  & BLEU	   &    0.243  & RougeL	& 0.365\\
XGB     	&	\textbf{0.434}  & C-BERT &    0.158  & Jaccard	& 0.234\\
DL 	&	0.425  & Rouge1  &	0.334  &  ChrF	 & 0.317\\\hline\hline
\end{tabular}
\end{center}
\end{table}

\subsection{Assessing and Improving LLM Performance in Statistical Programming}
Established datasets can be highly valuable for advancing research aimed at improving the robustness of code generated by LLMs. \shortciteN{Songetal2025-coderating} highlighted the weaknesses in LLM-generated statistical code, noting that it often lacks executability and produces inaccurate outputs. Thus, significant room for improvement remains. In this section, we explore how StatLLM can be leveraged to enhance the performance of LLMs in statistical programming.

The StatLLM dataset includes multiple LLM-generated code versions for each problem, based on both the problem description and data description. These diverse code variations offer an opportunity to create ensemble-based solutions for improved robustness and performance. Additionally, the StatLLM dataset enables the evaluation and comparison of different ensemble modeling techniques (\shortciteNP{xue2024multi} and \shortciteNP{shaikh2024fundamental}). Such investigations may lead to the development of new ensemble methods specifically designed for programming code data.

Moreover, the established datasets can facilitate new studies on evaluating LLM performance in generating programming code beyond SAS. Since the problem descriptions and data descriptions in the StatLLM dataset correspond to general statistical analysis tasks, the dataset can be extended to generate code in other languages, such as Python and R. This expansion enables investigations into LLM performance across different programming languages.

In addition to LLM-generated SAS code, StatLLM can be applied to other statistical programming languages, such as R, Python, and MATLAB. This flexibility encourages exploring language-specific challenges, examining cross-language robustness, and developing ensemble methods that integrate code generated from multi-programming languages to boost overall performance (\shortciteNP{xue2024multi}).

\subsection{Developing and Testing of Next-generation Statistical Software}
StatLLM can facilitate the development and testing of next-generation statistical software. The way humans communicate with statistical software has evolved (\shortciteNP{Minetal2024-appliedstat}), transitioning from command-line interfaces and compiled languages to graphical user interfaces (GUIs) with menus and clicks. These advancements have made statistical tools more accessible to non-statisticians for data analysis. By leveraging LLMs, next-generation statistical software will enable users to interact with the software using natural language.

\begin{figure}
\begin{center}
\includegraphics[width=0.75\textwidth]{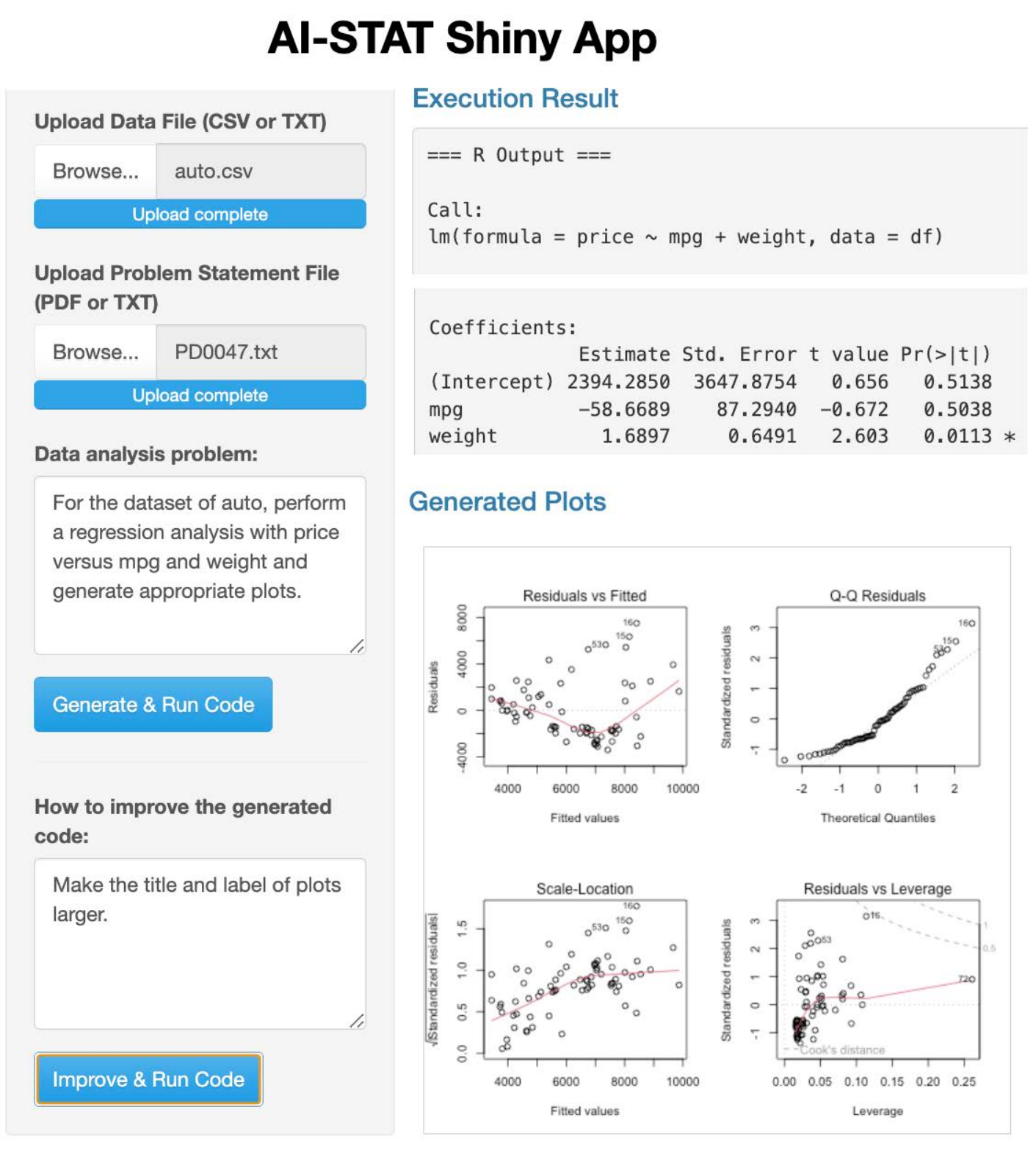}
\caption{An illustration of R Shiny app for showcasing automatic statistical analysis powered by AI technology.}\label{fig:rshiny}
\end{center}
\end{figure}

As an example, we developed an R Shiny app to demonstrate the potential of StatLLM in enabling automatic statistical analysis in R through R Shiny (\shortciteNP{shiny}). Figure~\ref{fig:rshiny} showcases an AI-powered automatic statistical analysis. As a preliminary step, this app leverages LLMs for automatic statistical analysis. When users upload their dataset along with a corresponding problem statement, the app forwards this information to ChatGPT 4.0, which generates R code for tasks ranging from data preprocessing and model fitting to visualization. The app then executes the generated code in R and displays the results through an interactive user interface. Additionally, the system supports iterative code refinement, allowing users to specify modifications for further customization. By streamlining this process, the app eliminates the need for manual R programming and running, significantly simplifying statistical analysis.

To evaluate the app's performance, we tested it using the \texttt{auto} dataset and a problem description from StatLLM. The dataset includes automotive specifications and pricing for various car models. Specifically, we instructed the app to perform a regression analysis, modeling car price as a function of miles per gallon (mpg) and weight. The app automatically generated and executed R code to fit a linear regression model, displaying both the model fitting results and diagnostic plots.

\section{Ethics and Fairness}\label{sec:ethics.fairness}

The StatLLM data collection process in this work does not involve individual privacy concerns. To the best of our knowledge, we have ensured that the data remains transparent and accountable. To enhance the fairness and reliability of rating score collection, we have fully incorporated design of experiments strategies, including randomization, blocking, and replication (\shortciteNP{wu2011experiments}). The questions and datasets used for analysis are sourced from publicly available materials, with a focus on statistical analysis tasks at the undergraduate and master's levels.

Our data collection procedure carefully applies randomization and ensures that rater identities remain anonymous during the scoring of LLM-generated code. To minimize potential biases, we employ both treatment randomization and randomized assignment when distributing code for evaluation (i.e., obtain the rating score). Additionally, LLM identities are concealed during grading to prevent biases related to model attribution.

To further control for bias, the sequence of problem descriptions presented to graders is randomized, reducing potential order and time effects. Moreover, each problem description and its corresponding LLM-generated code are evaluated by three independent graders independently, helping to reduce individual scoring variability and enhance the reliability of the ratings.

\section{Conclusions}\label{sec:conclusions}

In this work, we establish an open-source dataset to evaluate the performance of LLMs in statistical analysis. The dataset, named StatLLM, serves as a benchmark for AI assurance, particularly in AI-assisted statistical coding and AI-automated coding. StatLLM consists of three key components: statistical analysis tasks, LLM-generated SAS code, and human evaluation scores. These components provide valuable resources for addressing research gaps and enabling new studies, as illustrated in Section~\ref{sec:illustration.impact}. Additionally, the dataset enhances the reproducibility of AI-driven statistical methods across various applications.

Furthermore, StatLLM is highly extendable, allowing for the inclusion of additional components along multiple dimensions. Future expansions may incorporate more advanced statistical analysis tasks, LLM-generated code in other statistical programming languages, such as R, and alternative evaluation metrics. By continuously evolving, StatLLM can support broader research efforts and further advance the integration of AI in statistical analysis.

\section*{Acknowledgments}

The work by Hong and Deng was supported in part by the COS Dean's Discovery Fund (Award: 452021) and the work by Hong was supported in part by the Data Science Faculty Fellowship (Award: 452118) at Virginia Tech.

\appendix

\section{Online Data Repository}\label{sec:online.data.repo}

The dataset is available at \url{https://github.com/yili-hong/StatLLM}. The repository is systematically structured into three main directories: \path{Statistical_Analysis_Tasks}, \path{LLM_Generated_SAS_Code}, and \path{Human_Evaluation_Scores}, each containing well-organized subdirectories to facilitate efficient access to statistical problems, datasets, and SAS code.

\subsubsection*{1. Statistical Analysis Tasks}
This directory contains all essential resources required for statistical task solving. The file \path{Summary.csv} serves as a metadata file mapping each statistical task to corresponding DataDescription, DataSet, SAS code solutions and SAS code for data reading.

\begin{itemize}
 \item \path{ProblemDescription}: Provides detailed statistical task statements.
 \item \path{DataDescription}: Stores detailed dataset descriptions, including dataset name and variable definitions.
 \item \path{TaskDatasets}: Includes raw dataset files in CSV or XLSX format.
 \item \path{HumanVerified_SAScode}: Contains manually verified SAS solutions for each task.
 \item \path{DataReading_SASCode}: Contains SAS scripts designed to load and preprocess datasets.
\end{itemize}

\subsubsection*{2. LLM-Generated SAS Code}
This directory stores SAS code generated by different large language models (LLMs): GPT35, GPT4, and Llama. Each folder contains SAS scripts generated by the respective model under folder \path{SAS_code_only}.

\subsubsection*{3. Human Evaluation Scores}
This directory contains the three group scores, which are \path{CodeQuality_Score}, \path{CodeExecuta} \path{bility_Score}, and \path{CodeOutput_Score}, along with the \path{Total_score} for the three LLMs, evaluated across 207 statistical tasks. These scores are derived based on the SAS code found in the \path{LLM-generated SAS Code/SAS_code_only} folder. The corresponding task IDs are documented in \path{Summary.csv}.

To obtain the SAS results, users must run the corresponding \path{DataReading_SASCode} file along with either the \path{Humanverified_SAScode} or the \path{SAS_code_only} file from the LLM-generated outputs (refer to \path{Summary.csv}). For example, if solving the problem PD0007 with dataset DS0003, the user should first run \path{DataReading_SASCode/DR0003.txt} to properly import the dataset into SAS. After successfully loading the data, the user can then execute either \path{Humanverified_SAScode/SC0007.txt} (if using the manually verified solution) or \path{LLM_Generated_SAS_Code/GPT4/SAS_code_only/sas_query7.txt} (if testing the LLM-generated code).

\section{Calculation of NLP Metrics}\label{sec:calc.nlp.metrics}
To make the paper self-contain, in this appendix, we provide a brief introduction on the calculation of NLP metrics used in this paper.

\subsection{BLEU}

BLEU (Bilingual Evaluation Understudy) is the most commonly used metric in machine translation (\shortciteNP{papineni-etal-2002-bleu}). It measures the overlap of contiguous $n$-token sequences between machine-generated and human-referenced texts and ranges from 0 to 1, with higher scores indicating greater similarity. In this paper, BLEU has been adapted to assess the similarity between LLM-generated and human-verified SAS code.  Before computing $n$-gram overlaps, the SAS code is tokenized into elements such as PROC procedures, variables, and models. In particular, BLEU is computed as,
\begin{equation}
	\text{BLEU} = \text{BP} \times \exp\left( \sum_{n=1}^{N} w_n \log(p_n) \right),
\end{equation}
where the brevity penalty (BP) is
\begin{equation}\label{eq:BLEU_BP}
	\text{BP} =
	\begin{cases}
		1, & \text{if } c > r \\
		\exp(1-\frac{r}{c}), & \text{if } c \leq r
	\end{cases},
\end{equation}
where $c$ is length of the LLM-generated SAS code, $r$ is length of human-verified SAS code, $N$ is the maximum $n$-gram order (e.g., 4), $w_{n}$ is the weight for $n$-gram of order $n$ (e.g.,$\frac{1}{N}$), and $p_{n}$ is modified precision for $n$-grams as in,
\begin{equation}
	p_n = \frac{\sum_{g \in G} \min\{ \text{Count}_G(g), \text{MaxRefCount}(g)\}}{\sum_{g \in G} \text{Count}_G(g)},
\end{equation}
where $G$ is the set of $n$-grams, $\text{Count}_G(g)$ is the frequency of $g$ the LLM-generated SAS code. $\text{MaxRefCount}(g)$ is the maximum frequency of $g$ across the human-verified SAS codes.

\subsection{ROUGE}
ROUGE (Recall-Oriented Understudy for Gisting Evaluation) is a series of metrics (\shortciteNP{Lin2004ROUGEAP}) that can be used to measure the quality of LLM-generated SAS code, denoted by $L$, by comparing it against the human-verified SAS code, denoted by $H$. In this paper, we consider ROUGE-1, ROUGE-2, and ROUGE-L.

ROUGE-1 measures the overlap of unigrams (single words or tokens) between $H$ and $L$. we define,
\begin{equation}
	\text{R1 Recall} = \frac{\sum\limits_{w \in H} \min\{\text{Count}(w, L), \text{Count}(w, H)\}}{\sum\limits_{w \in H} \text{Count}(w, H)},
\end{equation}
\begin{equation}
	\text{R1 Precision} = \frac{\sum\limits_{w \in H} \min\{\text{Count}(w, L), \text{Count}(w, H)\}}{\sum\limits_{w \in L} \text{Count}(w, L)},
\end{equation}
where $\text{Count}(w, X)$ represents the number of times uni-gram $w$ happens at SAS code $X$. The the ROUGE-1 F1 score is computed as,
\begin{equation}
	\text{ROUGE-1 F1} = \frac{2 \times \text{R1 Precision} \times \text{R1 Recall}}{\text{R1 Precision} + \text{R1 Recall}}.
\end{equation}
The computing of ROUGE-2 is similar, which measures the overlap of bi-grams (consecutive word pairs) between $L$ and $H$.

ROUGE-L measures the longest common subsequence (LCS) between $H$ and $L$, which is computed as,
\begin{equation}
	\text{LCS}(H, L) = \max_{\substack{S}} | \text{LCS}(H, L) |,
\end{equation}
where $S$ denotes all common subsequences. We compute,
\begin{equation}
	\text{RL Recall} = \frac{\text{LCS}(H, L)}{|H|},
\end{equation}
\begin{equation}
	\text{RL Precision} = \frac{\text{LCS}(H, L)}{|L|},
\end{equation}
where $|H|$ and $|L|$ represent the number of tokens in  $H$ and $L$. The ROUGE-L F1 score is computed as,
\begin{equation}
	\text{ROUGE-L F1} = \frac{(1 + \beta^2) \times \text{RL Precision} \times \text{RL Recall}}{\beta^2 \times \text{RL Precision} + \text{RL Recall}},
\end{equation}
and the default setting of $\beta$ is 1.

\subsection{METEOR}
Meteor (Metric for Evaluation of Translation with Explicit ORdering) (\shortciteNP{banerjee-lavie-2005-meteor}) incorporates both precision and recall of uni-gram matches between the LLM-generated SAS code and the human-verified SAS code, and it is then adjusted by a fragmentation penalty. The Meteor is computed as,
\begin{equation}
	\text{METEOR} = F_{\text{mean}} \times (1 - p),
\end{equation}
\begin{equation}
	p = \gamma \left( \frac{c}{m} \right)^\beta, \label{eq:frag_penalty}
\end{equation}
where $p$ is fragmentation penalty, $F_{\text{mean}}$ is harmonic mean of recall and precision, $c$ is the number of chunks, $m$ is the umber of matched uni-grams between the LLM-generated SAS code and human-verified SAS code, and $\gamma$ and $\beta$ controls the magnitude of penalty.

\subsection{CodeBERTScore}

BERTScore leverages embeddings from pre-trained models like BERT to assess the semantic similarity of generated text (\shortciteNP{zhang2020bertscoreevaluatingtextgeneration}). CodeBERTScore is an advanced evaluation metric derived from BERTScore, specifically designed for code evaluation (\shortciteNP{zhou-etal-2023-codebertscore}). Given the encoded tokens for human-verified SAS code with mask, and the encoded tokens for LLM-generated SAS code with mask, we compute $\text{CB}_{\text{Precision}}$ and and $\text{CB}_{\text{Recall}}$ as the precision and the recall of CodeBERTscore, respectively.  Then, we compute the CodeBERT score as follow,
\begin{equation}
\text{CB} =  \frac{10 \times \text{CB}_{\text{Precision}} \times \text{CB}_{\text{Recall}}}{9 \times \text{CB}_{\text{Precision}} + \text{CB}_{\text{Recall}}}.
\end{equation}

\subsection{Character F-score (chrF)}

The Character $n$-gram F-score (chrF) is an evaluation metric (\shortciteNP{Popovic2015chrFCN}) that can be used to measure the similarity between LLM-generated SAS code, denoted by $L$, and human-verified SAS code, denoted by $H$.  The common $n$-gram count $C_{n}$ between $H$ and $L$ is:
\begin{equation}
C_n = \sum_{\gamma \in \mathcal{N}_n(L)} \min\{f(\gamma, L), f(\gamma, H)\},
\end{equation}
where $f(\gamma, L)$ indicates the frequency of an $n$-gram $\gamma$ in text $L$ and same for $f(\gamma, H)$, and $\mathcal{N}_n(L)$ denotes the set of character  $n$-grams extracted from text $L$.
The total candidate $n$-grams in the LLM-generated SAS code denoted as $T_{L}$ is defined as follows:
\begin{equation}
T_{L} = \sum_{\gamma \in \mathcal{N}_n(L)} f(\gamma, L).
\end{equation}
The total reference $n$-grams in the human-verified SAS code denoted as $T_{H}$ is defined as follows:
\begin{equation}
T_{H} = \sum_{\gamma \in \mathcal{N}_n(H)} f(\gamma, H).
\end{equation}
The precision and recall  are computed as:
\begin{equation}
P_n = \frac{C_n}{T_L}, \quad \text{ and }\quad
R_n = \frac{C_n}{T_H},
\end{equation}
respectively. The character $n$-gram F-score (chrF) is computed as:
\begin{equation}
\text{chrF}_n = (1 + \beta^2) \cdot \frac{P_n \cdot R_n}{\beta^2 \cdot P_n + R_n},
\end{equation}
where $\beta$ is a weighting factor: $\beta > 1$ gives more importance to recall and $\beta<1$ gives more weight to precision.

\subsection{Jaccard Similarity}

Jaccard similarity measures the overlap between the tokens (\shortciteNP{RinJaccard8089237}). It can be used to compare human-verified SAS code, denoted by $H$, and LLM-generated SAS code, denoted by $L$. The Jaccard Similarity is computed as,
\begin{equation}
	J(H, L) = \frac{|H \cap L|}{|H\cup L|},
\end{equation}
where $|H \cap L|$ represents the number of shared tokens in $H$ and $L$, and $ |H \cup L|$ represents the total number of unique elements in both sets. It is easy to see that the value of Jaccard Similarity ranges from 0 to 1, where 1 indicates that the two SAS scripts are identical while 0 indicates that there are no common words or tokens between the scripts, and values between 0 and 1 indicate partial similarity.


\end{document}